**LEAVE THIS PAGE BLANK
DO NOT DELETE THIS PAGE**

# BIM LOD + Virtual Reality

Using Game Engine for Visualization in Architectural & Construction Education

**Hassan Anifowose**
Texas A&M University
**Wei Yan**
Texas A&M University
**Manish Dixit**
Texas A&M University

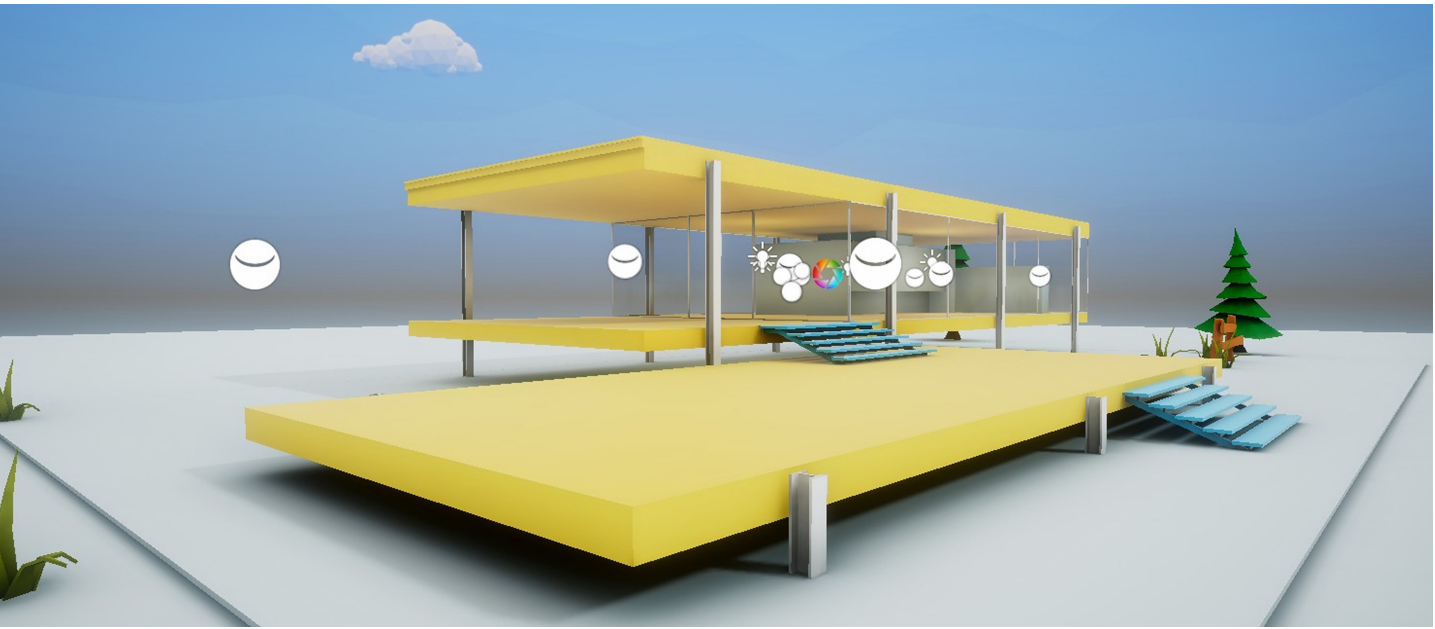

## ABSTRACT

Architectural Education faces limitations due to its tactile approach to learning in classrooms with only 2-D and 3-D tools. At a higher level, virtual reality provides a potential for delivering more information to individuals undergoing design learning. This paper investigates a hypothesis establishing grounds towards a new research in Building Information Modeling (BIM) and Virtual Reality (VR). The hypothesis is projected to determine best practices for content creation and tactile object virtual interaction, which potentially can improve learning in architectural & construction education with a less costly approach and ease of access to well-known buildings. We explored this hypothesis in a step-by-step game design demonstration in VR, by showcasing the exploration of the Farnsworth House and reproducing assemblage of the same with different game levels of difficulty which correspond with varying BIM levels of development (LODs). The game design prototype equally provides an entry way and learning style for users with or without a formal architectural or construction education seeking to understand design tectonics within diverse or cross-disciplinary study cases. This paper shows that developing geometric abstract concepts of design pedagogy, using varying LODs for game content and levels, while utilizing newly developed features such as snap-to-grid, snap-to-position and snap-to-angle to improve user engagement during assemblage may provide deeper learning objectives for architectural precedent study.

## INTRODUCTION

The integration of interactive building anatomy modeling into the construction education system improves learning while enabling effective knowledge transfer to learners (Park et al. 2016). The COVID-19 pandemic proved education's over-reliance on classroom to be inadequate and adaptation to future demands is overdue. Researchers indicated in previous works that interactive gaming improves learning and motivation (Kharvari and Höhl 2019) and it is also acknowledged that BIM provides the framework for developing 3D geometric data which can be used in the generation of levels and maps in games (Yan, Culp, and Graf 2011). Virtual Reality (VR) on the other hand is technology which is fast gaining adoption, however, with limitations based on extra work required to visualize BIM models (Zaker and Coloma 2018).

We used an empirical research method by developing a simple demonstration to test the hypothesis. The overall objective is to study how building anatomy modelling process combined with varying levels of development (LODs) help the learning process for architectural precedents. Evidence shows that using an anatomical approach, varying LODs provided in geometric form holds a potential to improve interaction and overall learning while studying architectural precedents. This demonstration shows the significance of new features for improved engagement in VR.

## STATE OF THE ART

VR provides full scale perception of spaces while accurately representing materiality in deeper levels of immersion when compared to desktop screens and projection systems (Angulo 2015). VR's common use is to enhance walking experiences inside and around a virtual structure, however, it has been used to evaluate form and design of a building (Abdelhameed 2013).

Increasing the feeling of presence has been determined to improve overall VR experiences, which must be accompanied with an increased level of realism within VR (Du et al. 2016). Outside environment realism, there are limitations of accessibility to data from BIM environments, requiring improvement (Kieferle and Woessner 2015). Our research towards gamification aims to help users create and manipulate objects within the VR environment thereby limiting dependence on the BIM environment for new objects generated in a game scene. Designing and providing content, interface design alongside coding and programming are some of the identified issues to overcome (Maghool, Moeini, and Arefazar 2018).

### VR + Architectural Education

The exploration of existing architectural projects, precedents or case studies is at the core of architectural education. Design students are generally expected to provide clarity of understanding of such precedents' abstract principles and tectonics before tendering design proposals during studio sessions. Visiting architectural precedents can become very costly for students and institutions depending on various factors. Additionally, there are no layers of interactivity involved in local visits, therefore, retention varies between students (Bourdakis and Charitos 1999; Kharvari and Höhl 2019).

Learning within Virtual Environments (VEs) can be developed to create a foundation for learning about Virtual environment design. Pushing learning in VE further within tutored environments, is proven to "benefit young architects on their first step towards understanding the essence of architecture" (Kamath, Dongale, and Kamat 2012). The speed of learning in architectural education has been attributed to students' ability to detect their errors faster and learn from them (Dvorak et al. 2005; Abdelhameed 2013). Moloney indicated that "The advantages of working in a real-time environment where early design iterations can be tested from multiple points of view" which has been established alongside limitations of communicating while wearing a VR device (Moloney and Harvey 2004; Milovanovic et al. 2017).

### VR + Construction Education

The study of building composition in construction education is limited by teaching strategies. The Interactive Building Anatomy Modeling (IBAM) research has indicated that an anatomical approach can deliver better learning outcomes in construction education via assemblage and gamified techniques (Park et al. 2016; P. Wang et al. 2018 ). This motivates our research idea for a virtual precedent study environment combined with BIM to provide an avenue where students apply their education to solve real problems. However, various settings for VEs require testing in order to determine which situation provides flexibility as well as improve virtual construction learning (Ghosh 2012). Gamifying construction learning is one method of continuously engaging the younger population (Generation Y) who have difficulties in engaging with traditional educational methods (Goedert et al. 2011).

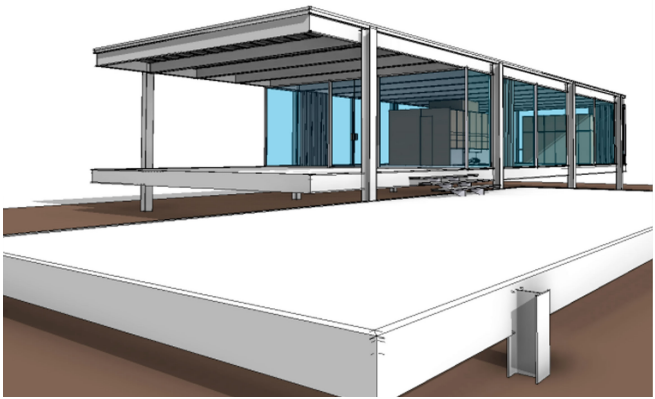

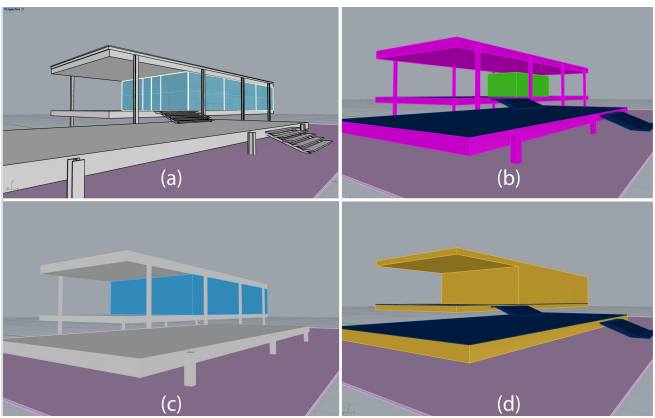

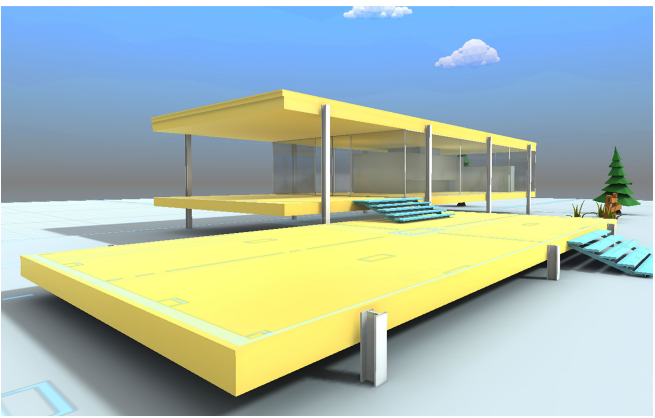

1   Completed detailed model of the building in Autodesk Revit.
2   Abstract Concept modeling of the Farnsworth House showing top left to bottom right, (a) full geometry, (b) structure, (c) lighting and (d) circulation respectively.
3   Precedent exploration in First Person View in the VR environment.
4   Stairs: LOD 200 (left) and LOD 300 (right).

The combination of building models in VR alongside a game process for learning, may hold considerable impact on learning when applied to education. As indicated in a research, learning in VR provides an experiential outcome which is obvious in the learner's behavior (R. Wang, Newton, and Lowe 2015).

## METHODS

### Gamification for Studying Architectural Precedents

Various classifications for Virtual Reality (VR) in Education have been established (Motejlek and Alpay 2019). These are - purpose of the application, technology of delivery, user interaction, user experience, system interaction and gamification types. Gamification is largely unexplored for developing learning experiences of Architectural precedents. Beyond the conventional methods, this paper showcases new techniques for studying architectural pedagogy through a new lens, i.e. Virtual Reality. To achieve cross-disciplinary, simplified learning for users without prior design knowledge, we utilized varying levels of development (LODs).

This study seeks answers to -

- What are best game design practices recommended to enhance playability in a BIM+VR game development?

### Pre-Game - Prototype Development

A prototype development provides an extensive tool for teaching varying LODs from concept design to final detailed design. Users explore a collaborative virtual environment and embark on design activities thereafter. For this study, we used the Farnsworth House.

Stage 1: Modeling– Detailed modelling of the building was completed in Autodesk Revit for studying (Figure 1). Abstract concepts are modeled in Rhinoceros and merged in Unity Game Engine using SteamVR plugin for gameplay and interactions.

The illustration (Figure 2), shows different concept schemes which provide learning about underlying design pedagogy.

Stage 2: Precedent Exploration – Users are introduced to the architectural precedent and allowed 10minutes for exploration (Figure 3).

Thereafter, users proceed to the game area for an assemblage exercise to test their knowledge.

Stage 3: Studio Setup, Practice Areas – In the VR environment, we developed a studio setup comprising an in-game projector screen which played a video narrating the

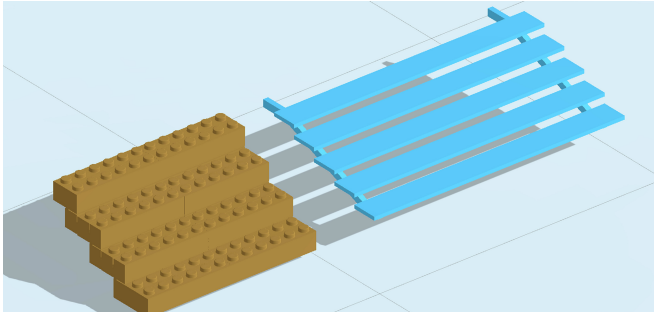

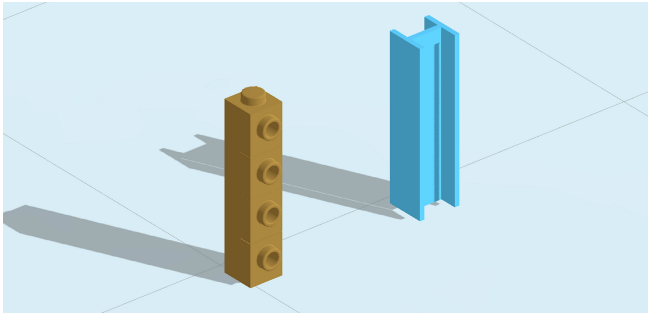

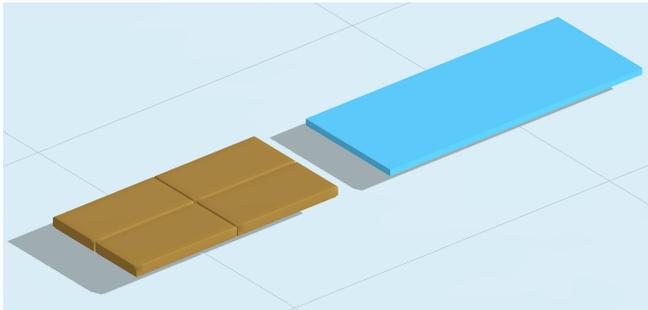

5  Steel Stanchion: LOD 200 (left) and LOD 300 (right).

6  Floor Slab: LOD 200 (left) and LOD 300 (right).

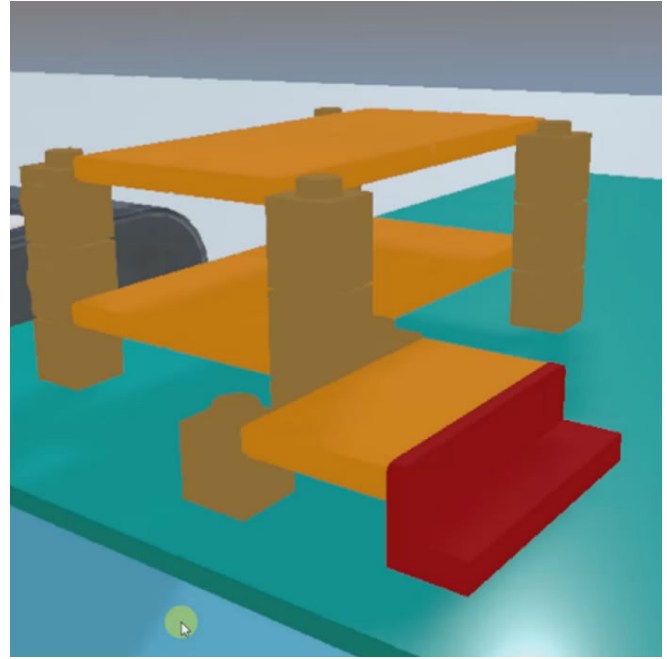

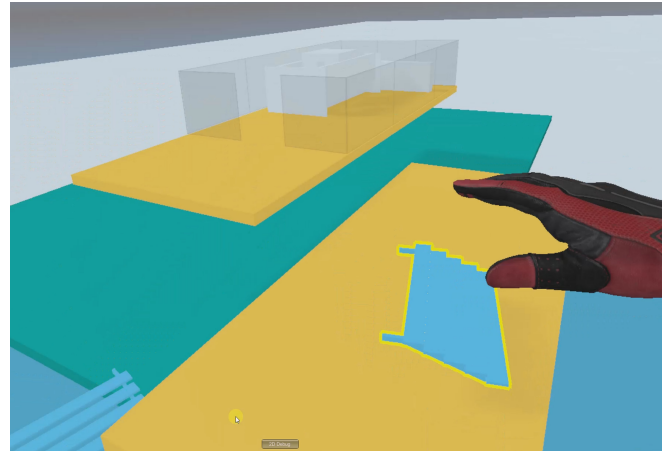

7  Completed Game Level 1 with LOD 200.

8  Game Challenge in progress for LOD 300.

architectural precedent while the game player explored the precedent and environment including interacting with abstract concept models in one case. The practice area in level 1 allows the player to get accustomed to object interaction and manipulation techniques required for the game challenge in level 2.

Stage 4: Gamification & Level Development – Borrowing from universally established BIM Level of Developments standard classifications (TrueCADD 2021), game levels were designed to correspond with increasing LOD. For the purpose of this research, we have classified the building construction components into game artifacts via two classes of LODs. The lowest LOD used is LOD 200 (approximate geometry represented by LEGO) while the highest LOD used is LOD 300 (precise geometry of the building). LEGO was used because of its geometric simplicity and popularity for low detail geometry assemblage. The classification is created specifically for this game based on geometry complexity and similar to the universal BIM LOD definition. Users are presented with each game level to complete assemblage tasks after the precedent explorations and pre-exercise stages. The assemblage task comprises of components with varying level of development (Figure 4, Figure 5, Figure 6).

Game level 1 features the player, assembling the building with LOD 200 construction components generated from Mecabricks web-based application with LEGO parts (Figure 7). Upon completion, the player proceeds to level 2 where they complete the assemblage with LOD 300 components (Figure 9) which are precise geometry.

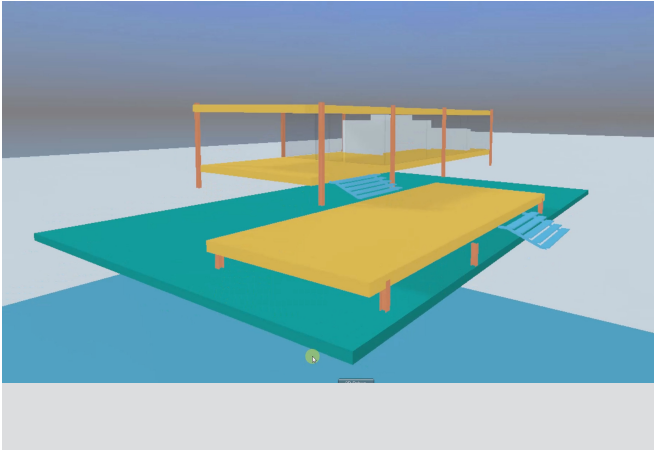

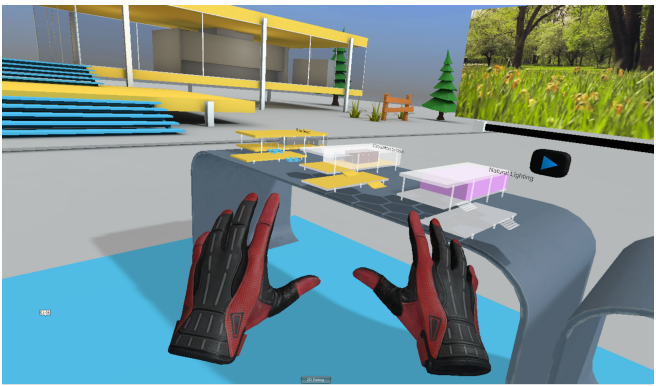

10

9  Completed Game Level 2 with LOD 300.

10  Game scene showing the base model and geometric abstract concepts (Circulation-to-use and Natural Lighting) in the exploration mode.

In-Game - Implementation and Prototype Testing
Our in-lab demonstration involved two players. We received immediate feedback on playability, implemented features and overall comfort within the VR environment. Player 1 explored the game with a studio environment setup, an in-game video narrative, and was presented with the assemblage task. Player 2 explored the game with all the features from Player 1's experience including interacting with the geometric abstract concepts (Figure 10). A new VR object-snapping feature developed for the game allowed both players to complete the tasks with the same levels of difficulty. Each player spent 10 minutes exploring the precedent and 10 minutes assembling the parts in the exercise area. The game was live demonstrated to a team of researchers at a college symposium. Demonstration video is available on https://youtu.be/_le_EpDeS9s

## RESULTS AND DISCUSSION

Feedback was collected from the two demonstration players based on four categories namely; player experience, user behavior, level of object interaction and knowledge retention. We observed that the game players were enthusiastic to learn about the architectural precedent. Player 2 had significant shorter time to reproduce the assemblage while also reporting a higher level understanding and motivation for learning and reproducing the building than Player 1. This indicates that having abstract concepts in precedent study virtual environments other than only the building geometry provides an opportunity for better knowledge retention. Players also indicated increased depth of understanding with the voice/video narration during the exploration stage. Especially important given the ongoing travel restrictions due to the pandemic, the BIM+VR prototype developed in this research using virtual game development modalities, provides a less costly approach and improved accessibility to well-known architectural precedents with a potential for learning and engagement. Institutions may adopt this strategy to provide rich learning objectives for both design-related and cross-disciplinary studies.

To improve user interaction and overall engagement, new features were created for the game which increased the intuition level for interacting with the assemblage components and increased the users' abilities to complete the assemblage tasks. By developing an object-spawning system (Figure 11) using the Bolt Visual Scripting tool, the dependence on importing heavy BIM files is drastically reduced thereby increasing adoption possibilities of this game's approach to learning in Virtual Reality. This tool was used in developing the embedded features saving time beyond the conventional C# scripting tools. This demonstration enabled us establish a case scenario for application within the design and construction education environment. As shown in (Figure 12), the BIM LOD + VR game research is grouped into four major stages of work. The first two stages and a portion of the third stage are reported in this paper.

## CONCLUSION

Virtual reality can potentially eliminate the need for local and international travel in architectural studies. This paper investigated the possibilities of using BIM+VR in a gamified LOD test case, as a way of increasing effective learning objectives in architectural forms and tectonics. Feedback from the demonstration indicated that higher understanding of the main architectural precedent could be gained by studying the developed abstract concepts in geometric form with varying LODs inside the game before attempting the assemblage task. In diverse cases where the user has no previous knowledge of architecture or construction, the BIM+VR approach provides learning

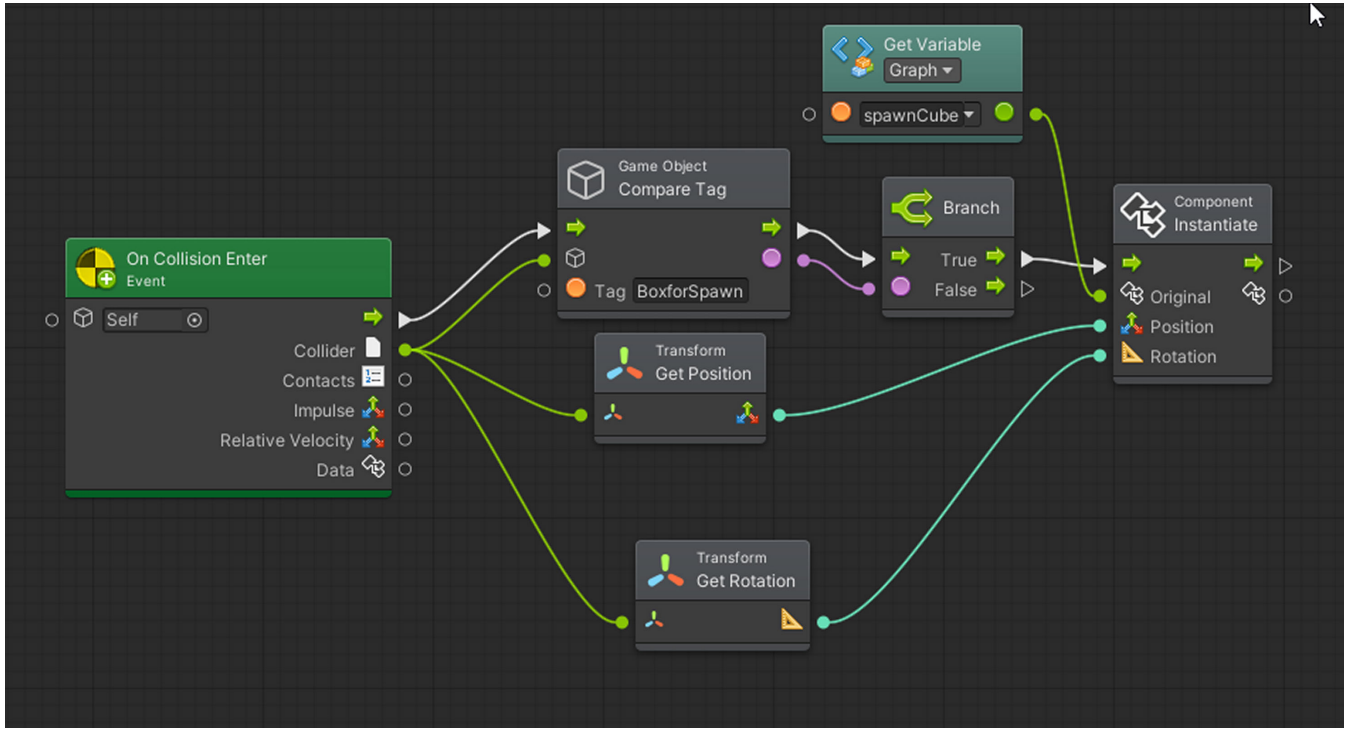

11  Object spawning system for game artifacts using Bolt Visual Scripting alternative beyond reading technical drawings; a task typically difficult for the majority in other fields of study. Developed features such as snap-to-grid, snap-to-angle and snap-to-position are also necessary to improve user engagement during the assemblage phase of the game. By generalizing results, we can confirm that this approach could provide a deeper knowledge of architectural precedents while simulating a design studio and modeling environment inside virtual reality. Limitations experienced included difficulty in generating custom scripts, and retention of BIM geometry data inside Unity software while converting objects from native Revit to game objects. More challenges were faced in developing mechanics for improved game performance. This was overcome by developing object spawning systems.

A larger scale experiment is currently being developed for construction education to enable researchers answer questions as to how well effective learning can occur using BIM and VR with varying LODs and also, determine which gamification features enhance interaction while contributing to effective learning. This will be carried out as a full study with a larger group of human subjects. With the findings in this research, gamification is expected to make its way into Architectural and Construction Education both in the classroom and on the field.

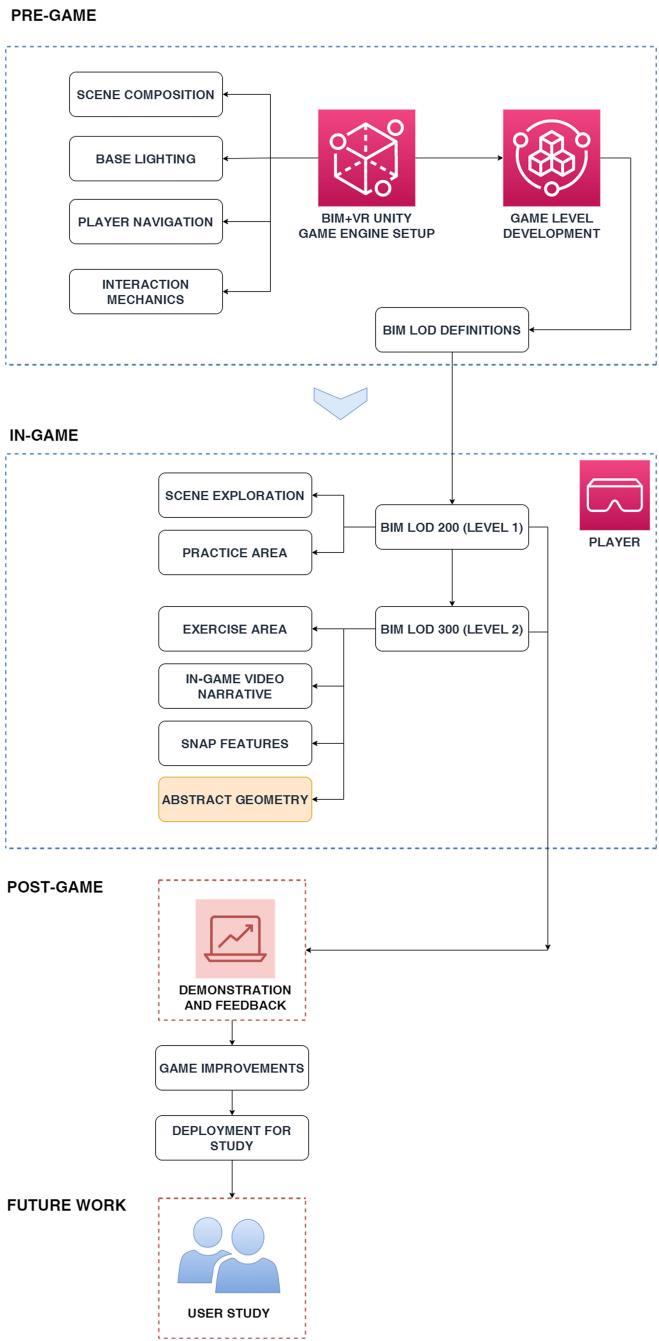

12  Diagram showing 4 stages of this research and features of each stage.

**Hassan Anifowose** An Architect and 3D Visualization expert with several years of experience in Design, Construction and Entrepreneurship. Hassan is a PhD Student at Texas A&M University where he is constantly working on the edge with interests spanning across a wide range of subjects including user experience, game design and human building interaction. He has won 8 scholarships during his PhD Program including the AIA's David Lakamp Scholarship. Hassan is presently exploring new frontiers of BIM and Virtual Reality applications for improved project performance on construction projects and aims to improve productivity in the construction industry with technology.

**Wei Yan** Dr. Wei Yan, Mattia Flabiano III AIA/Page Southerland Page Endowed Professor of Architecture at Texas A&M University, with expertise in Computational Methods in Design, Building Information Modeling, Augmented Reality, etc. He has led research projects funded by the National Science Foundation, the National Endowment for the Humanities, etc. He received the Best Paper Prize in Design Computing and Cognition '06. He was named a Texas A&M Presidential Impact Fellow in 2017. Yan was educated at the University of California, Berkeley (Ph.D. in Architecture and M.S. in Computer Science), ETH Zurich, and Tianjin University.

**Manish Dixit** PhD, LEED AP. Dr. Dixit is Associate Professor in Construction Science at Texas A&M University with over 10 years of experience in the design and commercial construction industry. His research interests include life-cycle energy and environmental modeling, green building materials, embodied energy modeling, 3-D printing in construction, Building Information Modeling (BIM) and facilities' performance assessment. He served as National Participant in Annex57 of the International Energy Agency (IEA) and endorsed as an expert in embodied energy analysis field by the US Department of Energy (USDOE). He represents the nation in Annex 72 (IEA) focusing on life cycle energy assessment and optimization of buildings.